\documentstyle[sprocl]{article}

\input{psfig}

\def\be{\begin{equation}}
\def\ee{\end{equation}}
\def\bea{\begin{eqnarray}}
\def\eea{\end{eqnarray}}

\begin{document}

\title{String Expansion as 't Hooft's Expansion \footnote{Report-no: HUTP-98/A053, NUB 3182}
\footnote{Talk presented at PASCOS'98} \footnote{This review is based on \cite{BKV,zura}.}}

\author{Zurab Kakushadze}

\address{Lyman Laboratory of Physics, Harvard University, Cambridge, 
MA 02138\\
and\\
Department of Physics, Northeastern University, Boston, MA 02115\\
E-mail: zurab@string.harvard.edu}

\maketitle\abstracts{We briefly review some recent developments in large $N$ gauge theories
which utilize the power of string perturbation techniques.}

\section{Introduction}

{}'t Hooft's large $N$ expansion \cite{thooft} is an attractive possibility
for understanding gauge theories. In this limit the gauge theory diagrams
look like Riemann surfaces with boundaries and handles. It is therefore
natural to attempt to map the large $N$ expansion of gauge theories
to some kind of string world-sheet expansion.

{}The first concrete example of such a map was given by Witten \cite{CS} for 
the case of three dimensional Chern-Simons gauge theory where the 
boundaries of the string world-sheet are ``topological'' D-branes. 

{}Recently the map between the large $N$ expansion and string expansion
has been made precise \cite{BKV} in the context of Type II string theory. 
The idea of \cite{BKV} is to consider Type IIB string theory with a large number 
$N\rightarrow\infty$ of D3-branes and take a limit where $\alpha^\prime\rightarrow 0$ while 
keeping $\lambda=N\lambda_s$ fixed, where $\lambda_s$ is the Type IIB string 
coupling. In this setup we have four dimensional 
gauge theories with unitary gauge groups. A 
world-sheet with $b$ boundaries and $g$ handles is weighted with
\begin{equation}\label{thoo}
 (N\lambda_s)^b \lambda^{2g-2}_s=\lambda^{2g-2+b} N^{-2g+2}~.
\end{equation} 
Upon identification $\lambda_s=g^2_{YM}$, this precisely maps to 't Hooft's
large $N$ expansion. This expansion is valid in the limit where $N\rightarrow\infty$
and the effective coupling $\lambda$ is fixed at a weak coupling value.

{}In \cite{BKV} the above idea was applied to prove that four dimensional 
gauge theories (including the cases with no supersymmetry) considered in
\cite{KaSi,LNV} 
are conformal to all orders in perturbation theory in the large $N$ 
limit. In particular, the corresponding gauge theories were obtained from 
Type IIB string theory with D3-branes embedded in orbifolded space-time.
The ultraviolet finiteness of string theory (that is, one-loop tadpole cancellation 
conditions) was shown to imply that the resulting (non-Abelian) gauge theories
where conformal in the large $N$ limit (in all loop orders). Moreover, in \cite{BKV}
it was also proven that computation of any correlation function in these theories 
in the large $N$ limit reduces to the corresponding computation in the parent 
${\cal N}=4$ supersymmetric gauge theory\footnote{For the field theory discussion, see \cite{BJ}.}. 

{}The all-order proofs in \cite{BKV} were possible due to the fact that the power
of string perturbation techniques was utilized. In particular, string theory perturbation
expansion is a very efficient way of summing up various field theory diagrams.
Thus, often a large number of field theory Feynman diagrams in a given order of
perturbation theory correspond to a single string theory diagram with certain
topology.
This has been successfully exploited to compute tree and loop level scattering 
processes in gauge theories \cite{Koso}. (For a recent discussion, see, {\em e.g.}, 
\cite{Dix}.) The all-order proofs in \cite{BKV} crucially depended on the fact that
the string world-sheet expansion was self-consistent. In particular, the arguments
in \cite{BKV} would not go through if the tadpoles were not cancelled. The tadpole
cancellation conditions, however, ultimately produced theories which were 
(super)conformal in the large $N$ limit. In particular, the one-loop $\beta$-functions 
(for non-Abelian gauge groups) in all of those theories were zero (even at finite $N$).

{}The setup of \cite{BKV} allows to consider gauge theories with unitary gauge groups. 
Moreover, the matter in these theories consists of bi-fundamentals/adjoints in the product gauge 
group. To include orthogonal and symplectic gauge groups (as well as other representations for the matter fields) we must consider 
Type IIB string theory with D3-branes as well as orientifold planes.
(In certain cases string consistency will also require presence of D7-branes.)
This leads us to consider Type IIB orientifolds where we 
can expect appearance of $SO$ and $Sp$ gauge groups. The generalization of \cite{BKV} to orientifolds was given in \cite{zura}. In fact, generically these orientifold theories have non-vanishing one-loop $\beta$-functions. However, the running of the gauge couplings is suppressed in the large $N$ limit (so in this sense these theories are ``finite'' in the large $N$ limit).

{}Introduction of orientifold planes changes the possible topologies of the world-sheet.
Now we can have a world-sheet with $b$ boundaries (corresponding to D-branes), 
$c$ cross-caps (corresponding to orientifold planes), and $g$ handles. Such a 
world-sheet is weighted with     
\begin{equation}\label{thoo1}
 (N\lambda_s)^b \lambda_s^c \lambda^{2g-2}_s=\lambda^{2g-2+b+c} N^{-c-2g+2}~.
\end{equation}
Note that addition of a cross-cap results in a diagram suppressed by an additional 
power of $N$, so that in the large $N$ limit we can hope for simplifications (or, rather,
we can hope to avoid complications with unoriented world-sheets, at least in some 
cases). In fact, for string vacua which are perturbatively consistent 
(that is, the tadpoles cancel) calculations of correlation functions 
in ${\cal N}<4$ gauge theories
reduce to the corresponding calculations in the parent ${\cal N}=4$
{\em oriented} theory. This holds not only for finite (in the large $N$ limit) gauge theories
but also for the gauge theories which are not conformal. 

{}It is very satisfying to observe that in the large $N$ limit
using the power of string theory perturbation techniques
we can reduce very non-trivial calculations in gauge theories with lower supersymmetries
to calculations in ${\cal N}=4$ gauge theories. In particular, this applies to
multi-point correlators in gauge theory. Moreover, these statements hold even in non-supersymmetric cases.

{}Here we note that the correspondence between 't Hooft's large $N$ expansion and string 
world-sheet expansion is expected to hold only in the regime where the effective
coupling $\lambda$ is small. If $\lambda$ is large one expects an effective supergravity description to take over 
\cite{Mald}.
The supergravity picture has, in particular, led to the conjectures 
in \cite{KaSi} as well as in \cite{LNV} about 
finiteness of certain gauge theories. However, proofs of those conjectures 
(in the large $N$ limit)
presented in \cite{BKV} were given in the {\em weakly} coupled region. Also, $1/N$ corrections
can only be reliably computed in this region but not in the strong coupling regime where
{\em a priori} there is no world-sheet expansion nor 't Hooft's expansion is valid.

{}The remainder of this talk is organized as follows. In section \ref{finite} we review the
arguments of \cite{BKV} for the cases without the orientifold planes. In section \ref{finite1} we
review generalization of these arguments 
to the orientifold cases discussed in \cite{zura}. 
In section 4 we discuss various issues relevant for the previous discussions. 

\section{Large $N$ Limit and Finiteness}\label{finite}

{}In this section we review the discussion in \cite{BKV} for the cases without orientifold
planes. We will discuss generalization \cite{zura} of these arguments to the orientifold cases in section \ref{finite1}.

\subsection{Setup}

{}Consider Type IIB string theory with $N$ parallel D3-branes where
the space transverse to the
D-branes is ${\cal M}={\bf R}^6/\Gamma$.
The orbifold group
$\Gamma= \left\{ g_a \right\}$ ($g_1=1$, $a=1,   \dots,
|\Gamma|$)
must be a finite discrete subgroup\footnote{In the following we will confine our attention
to orbifolds without discrete torsion. Cases with discrete torsion have recently been discussed
in \cite{doug}.} of $Spin(6)$.
If $\Gamma\subset SU(3)$ ($SU(2)$), we have
${\cal N}=1$ (${\cal N}=2$) unbroken supersymmetry,
and ${\cal N}=0$, otherwise.

{}Let us confine our attention to the cases where type IIB on ${\cal M}$ is
a modular invariant theory\footnote{This is always the case if $\Gamma\subset SU(3)$.
For the non-supersymmetric cases this is also true provided that
$\not\exists{\bf Z}_2\subset\Gamma$. If $\exists{\bf Z}_2\subset\Gamma$,
then modular invariance requires that the set of points in ${\bf R}^6$
fixed under the ${\bf Z}_2$ twist has real dimension 2.}. The action of the
orbifold on
the coordinates $X_i$ ($i=1,\dots,6$) on ${\cal M}$ can be described
in terms of $SO(6)$ matrices:
$g_a:X_i\rightarrow \sum_j (g_a)_{ij} X_j$.
We need to specify
the action of the orbifold group on the Chan-Paton charges carried by the
D3-branes. It is described by $N\times N$ matrices $\gamma_a$ that
form a representation of $\Gamma$. Note that $\gamma_1$ is an identity
matrix and ${\mbox {Tr}}(\gamma_1)=N$.

{}The D-brane sector of the theory is described by an {\it oriented} open
string theory. In particular, the world-sheet expansion corresponds
to summing over oriented Riemann surfaces with arbitrary genus $g$ and
arbitrary number of boundaries $b$, where the boundaries of the world-sheet 
correspond to the D3-branes.

{}For example, consider one-loop vacuum amplitude ($g=0$, $b=2$). The
corresponding graph is an annulus whose boundaries lie on D3-branes.
The one-loop partition function in the
light-cone gauge is given by
\begin{equation}\label{partition}
 {\cal Z}={1\over 2|\Gamma|}\sum_a
 {\rm Tr}  \left( g_a (1+(-1)^F)
 e^{-2\pi tL_0}
 \right)~,
\end{equation}
where $F$ is the fermion number operator, $t$ is the real modular parameter
of the annulus, and the trace includes sum over the Chan-Paton factors.

{}The orbifold group $\Gamma$ acts on both ends of the open strings.
The action of $g_a\in \Gamma$ on Chan-Paton charges is given by
$\gamma_a\otimes \gamma_a$. Therefore,
the individual terms in the sum in (\ref{partition})
have the following form:
\begin{equation}
 \left({\mbox {Tr}}(\gamma_a)\right)^2 {\cal Z}_a~,
\end{equation}
where ${\cal Z}_a$ are characters
corresponding to the world-sheet degrees of freedom. The ``untwisted''
character
${\cal Z}_1$ is the same as in the ${\cal N}=4$ theory for which
$\Gamma=\{1\}$. The information about the fact that the orbifold theory
has reduced supersymmetry is encoded in the ``twisted'' characters
${\cal Z}_a$, $a\not=1$.

{}In \cite{BKV} it was shown that the one-loop massless (and, in non-supersymmetric 
cases, tachyonic) tadpole cancellation conditions require that
\begin{equation}\label{pole}
 {\mbox {Tr}}(\gamma_a)=0~\forall a\not=1~.
\end{equation}
It was also shown that this condition implies that the Chan-Paton matrices $\gamma_a$
form an $n$-fold copy of the {\em regular} representation of $\Gamma$. The regular representation decomposes into a direct sum of all irreducible
representations ${\bf r}_i$ of $\Gamma$ with degeneracy factors
$n_i=|{\bf r}_i|$. The gauge group is ($N_i\equiv nn_i$)
\begin{equation}
 G=\otimes_i U(N_i)~. 
\end{equation}
The matter consists of Weyl fermions/scalars transforming in
bi-fundamentals
$({\bf N}_i,{\overline {\bf N}}_j)$ according to the decomposition
of the tensor product
of ${\bf 4}$ (${\bf 6}$) of $Spin(6)$ with the corresponding representation
(see \cite{LNV} for details).

\subsection{Large $N$ Limit}

{}The gauge group in the theories we are considering here
is $G=\otimes_i U(N_i) (\subset  U(N))$. In the following we will ignore the
$U(1)$ factors (for which the gauge couplings do run for ${\cal N}<4$ as there
are matter fields charged under them) and consider
$G=\otimes_i SU(N_i)$. In this subsection we review the arguments of
\cite{BKV} which show that in the large $N$
limit this non-Abelian gauge theory is 
conformal\footnote{Including the $U(1)$ factors does not alter the conclusions 
as their effect is subleading in the $1/N$ expansion \cite{BKV}.}.

{}There are two classes of diagrams we need to consider: ({\it i}) diagrams
without handles;
({\it ii}) diagrams with handles. The latter correspond to closed string loops
and are
subleading in the large $N$ limit. The diagrams without handles can 
be divided into two classes: ({\it i}) planar
diagrams
where all the external lines are attached to the same boundary; ({\it ii})
non-planar diagrams
where the external lines are attached to at least two different boundaries. The
latter are subleading in the large $N$ limit.

{}In the case of planar diagrams we have $b$ boundaries with all $M$ 
external lines attached to the same boundary (which without loss of generality
can be chosen to be the outer boundary) as depicted in Fig.1.
\begin{figure}[t]
\centerline{\psfig{figure=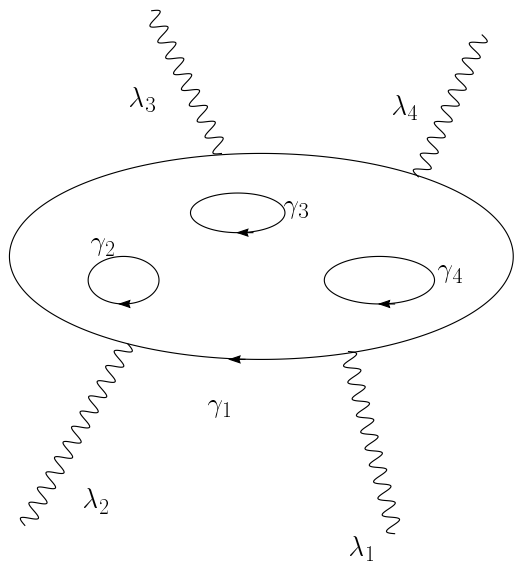,height=3 in}}
\caption{A planar diagram.}
\end{figure}
We need to sum over all possible twisted boundary conditions for the boundaries.
The boundary conditions must satisfy the requirement that
\begin{equation}\label{mono}
 \gamma_{a_1}=\prod_{s=2}^{b} \gamma_{a_s}~,
\end{equation}
where $\gamma_{a_1}$ corresponds to the outer boundary (to which we have attached
the external lines), and 
$\gamma_{a_s}$ ($s=2,\dots,b$) correspond to the inner boundaries (with no external lines).
Here we have chosen the convention (consistent with the corresponding choice made
for the annulus amplitude in (\ref{partition})) that the outer and inner boundaries have opposite
orientations. Then the above condition is simply the statement that only the states
invariant under the action of the orbifold group contribute into the amplitude.

{}If all the twisted boundary conditions are trivial ({\em i.e.}, $a_s=1$ for all $s=1,\dots,b$)
then the corresponding amplitude is the same as in the 
${\cal N} =4$ case (modulo factors of $1/{|\Gamma|}$ coming from 
the difference in normalizations of the corresponding D-brane boundary states
in the cases with ${\cal N}=4$ (where $|\Gamma|=1$) and ${\cal N}<4$ (where 
$|\Gamma|\not=1$)). Therefore, such amplitudes do not contribute to the gauge
coupling running (for which we would have $M=2$ gauge bosons attached to
the outer boundary) since the latter is not renormalized in ${\cal N}=4$ gauge 
theories due to supersymmetry. 

{}Let us now consider contributions with non-trivial twisted
boundary conditions. Let $\lambda_r$, $r=1\dots M$, be the Chan-Paton matrices corresponding to the
external lines. Then the planar diagram with $b$ boundaries has the following
Chan-Paton group-theoretic dependence:
\begin{equation}
 \sum {\rm Tr}\left(\gamma_{a_1} \lambda_1\dots\lambda_M\right)
\prod_{s=2}^{b}  {\rm Tr}(\gamma_{a_s})~,
\end{equation}
where the sum involves all possible distributions of $\gamma_{a_s}$ twists
(that satisfy the condition (\ref{mono})) as well as permutations of $\lambda_r$
factors (note that the $\lambda$'s here are the states
which are kept after the orbifold projection, and so they commute 
with the action of $\gamma$'s). The important point here
is that unless all twists $\gamma_{a_s}$ are trivial for $s=2,\dots,b$, the
above diagram vanishes by the virtue of
(\ref{pole}). But then from (\ref{mono}) it follows that $\gamma_{a_1}$
must be trivial as well. This implies that the only planar diagrams that
contribute are those with trivial boundary conditions which (up to numerical
factors) are the same as in the parent ${\cal N}=4$ gauge theory. 
This establishes that computation of any $M$-point correlation function in the
large $N$ limit reduces to the corresponding ${\cal N}=4$ calculation, and that
these gauge theories are (super)conformal in this limit\footnote{In \cite{BKV} it was also 
shown that a large class of non-planar diagrams without handles also vanish. We refer
the reader to \cite{BKV} for details.}.

{}Here we should mention that the models of \cite{BKV} are perturbatively
consistent string theories at all energy scales. In particular, the Abelian factors (that 
run in the low energy effective theory and decouple in the infrared) are not problematic
from the string theory viewpoint (although in the field theory context they would have
Landau poles in the ultraviolet).

\section{Generalization to Orientifolds}\label{finite1}

{}In this section we review generalization of the approach of \cite{BKV} to theories with
orientifold planes \cite{zura}.   

\subsection{Setup}

{}Consider Type IIB string theory on ${\cal M}={\bf C}^3/\Gamma$ where
$\Gamma\subset Spin(6)$. Consider the $\Omega J$ orientifold of this 
theory, where $\Omega$ is the world-sheet parity reversal, and $J$ 
is a ${\bf Z}_2$ element ($J^2=1$) acting on the complex coordinates $z_i$
($i=1,2,3$) on ${\bf C}^3$ such that the set of points in ${\bf C}^3$ fixed under 
the action of $J$ has real dimension $\Delta=0$ or $4$. 

{}If $\Delta=0$ then we have an orientifold 3-plane. If $\Gamma$ has
a ${\bf Z}_2$ subgroup, then we also have an orientifold 7-plane.
If $\Delta=4$ then we have an orientifold 7-plane. We may also have
an orientifold 3-plane depending on whether $\Gamma$ has an appropriate
${\bf Z}_2$ subgroup. Regardless of whether we have an orientifold 3-plane,
we can {\em a priori} introduce an arbitrary number of D3-branes (as the corresponding
tadpoles vanish due to the fact that the space transverse to the D3-branes
is non-compact). On the other hand, if we have an orientifold 7-plane we must 
introduce 8 of the corresponding D7-branes to cancel the R-R charge appropriately.
(The number 8 of D7-branes is required by the corresponding tadpole cancellation
conditions.) 

{}We need to specify the action of $\Gamma$ on the Chan-Paton factors
corresponding to the D3- and/or D7-branes. Just as in the previous section, 
these are given by Chan-Paton matrices which we collectively refer to
as $\gamma^\mu_a$, where the superscript $\mu$ refers to the corresponding
D3- or D7-branes. Note that ${\mbox{Tr}}(\gamma^\mu_1)=n^\mu$ where 
$n^\mu$ is the number of D-branes labelled by $\mu$. 

{}At one-loop level there are three different sources for massless tadpoles:
the Klein bottle, annulus, and M{\"o}bius strip amplitudes depicted in Fig.2. 
The Klein bottle amplitude corresponds to the contribution of unoriented 
closed strings into one-loop vacuum diagram. It can be alternatively viewed
as a tree-level closed string amplitude where the closed strings propagate 
between two cross-caps. The latter are (coherent Type IIB) states 
that describe the familiar orientifold planes. The annulus amplitude 
corresponds to the contribution of open strings stretched between two D-branes
into one-loop vacuum amplitude. It can also be viewed as a tree-channel closed
string amplitude where the closed strings propagate between two D-branes.
Finally, the M{\"o}bius strip amplitude corresponds to the contribution of unoriented
open strings into one-loop vacuum diagram. It can be viewed as a tree-channel closed 
string amplitude where the closed strings propagate between a D-brane and an orientifold
plane.

\begin{figure}[t]
\centerline{\psfig{figure=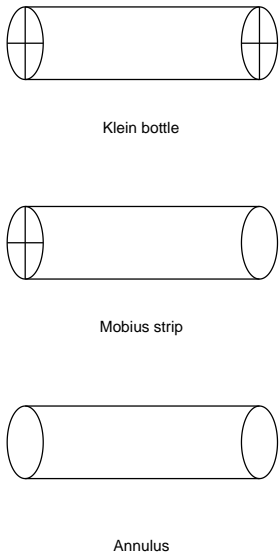,height=3 in}}
\caption{The Klein bottle, M{\"o}bius strip and annulus amplitudes.}
\end{figure}

{}Note that there are no Chan-Paton matrices associated with the Klein bottle 
amplitude since it corresponds to closed strings propagating between two cross-caps
which do not carry Chan-Paton charges. The M{\"o}bius strip has only one boundary.
This implies that the individual terms (corresponding to twists $g_a\in \Gamma$)
in the M{\"o}bius strip amplitude are proportional to ${\mbox{Tr}}(\gamma^\mu_a)$. The annulus
amplitude is the same (up to an overall factor of $1/2$ due to the orientation reversal projection) 
as in the oriented case discussed in the previous section. Thus, the individual terms (corresponding to twists $g_a\in \Gamma$)
in the annulus amplitude are proportional to ${\mbox{Tr}}(\gamma^\mu_a){\mbox{Tr}}(\gamma^\nu_a)$. 
Thus, the tadpoles can be written as

\begin{figure}[t]
\centerline{\psfig{figure=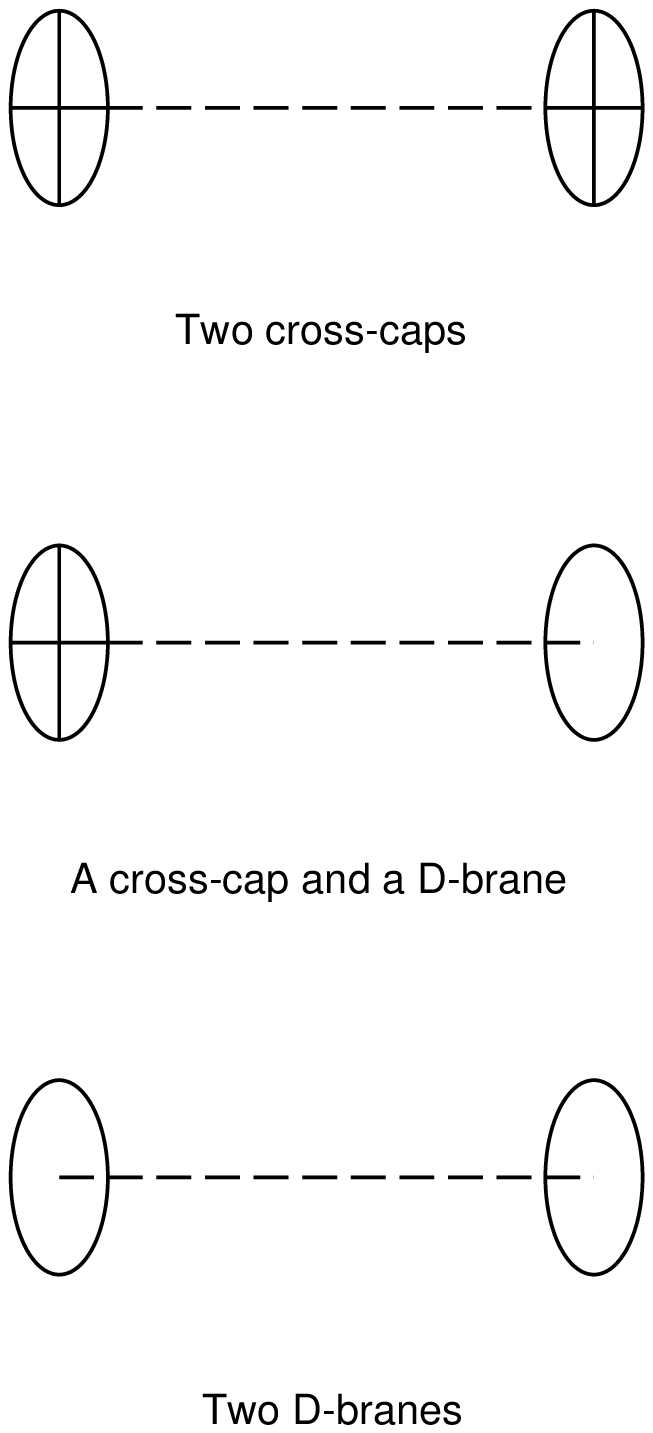,height=3 in}}
\caption{Factorization of the Klein bottle, M{\"o}bius strip and annulus amplitudes.}
\end{figure}

\begin{equation}\label{KMA}
 \sum_a \left( K_a +\sum_\mu M^\mu_a {\mbox{Tr}}(\gamma^\mu_a)+
                         \sum_{\mu,\nu}A^{\mu\nu}_a
 {\mbox{Tr}}(\gamma^\mu_a){\mbox{Tr}}(\gamma^\nu_a)
 \right)~.
\end{equation}   
Here terms with $K_a$, $M^\mu_a$ and $A^{\mu\nu}_a$ correspond to the
contributions of the Klein bottle, M{\"o}bius strip and annulus amplitudes, respectively.
In fact, the factorization property of string theory implies that the Klein bottle amplitude
should factorize into two cross-caps connected via a long thin tube. The M{\"o}bius strip
amplitude should factorize into a cross-cap and a disc connected via a long thin tube.
Similarly, the annulus amplitude should factorize into two discs connected via a long thin tube.
These factorizations are depicted in Fig.3. 
The implication of this for the tadpoles is that
they too factorize into a sum of perfect squares
\begin{equation}\label{tad}
 \sum_a \left(B_a+\sum_\mu C^\mu_a {\mbox{Tr}}(\gamma^\mu_a)\right)^2~,
\end{equation}  
where $B_a^2=K_a$, $2B_a C^\mu_a=M^\mu_a$ and $C^\mu_a C^\nu_a=A^{\mu\nu}_a$.
Thus, the tadpole cancellation conditions read:
\begin{equation}
 B_a+\sum_\mu C^\mu_a {\mbox{Tr}}(\gamma^\mu_a)=0~.
\end{equation}
Note that 
\begin{equation}\label{Klein}
 {\mbox {Tr}}(\gamma^\mu_a)=0~\forall a\not=1~{\mbox{only if}}~K_a=0~\forall a\not=1~.
\end{equation}
If this condition is satisfied then the
corresponding (non-Abelian) gauge theories are superconformal in the large $N$ limit
\cite{zura}.
On the other hand, if not all $K_a$ with $a\not=1$ are zero, then some of the Chan-Paton
matrices $\gamma^\mu_a$ with $a\not=1$ must have non-zero traces. This will
generically lead to theories with non-vanishing one-loop $\beta$-functions.

\subsection{Large $N$ Limit}

{}In this subsection we extend the arguments reviewed in section \ref{finite}
to the cases with orientifold planes. In particular, we will study the large $N$ behavior
of $M$-point correlators of fields charged under the gauge group that arises
from the D3-branes. In the following we will ignore the
$U(1)$ factors (if any) in the D3-brane gauge group.

{}There are two classes of diagrams we need to consider: ({\it i}) diagrams
without handles and cross-caps;
({\it ii}) diagrams with handles and/or cross-caps. 
The latter are
subleading in the large $N$ limit. The diagrams without handles and cross-caps
can 
be divided into two classes: ({\it i}) planar
diagrams
where all the external lines are attached to the same boundary; ({\it ii})
non-planar diagrams
where the external lines are attached to at least two different boundaries. The
latter are subleading in the large $N$ limit.

{}In the case of planar diagrams we have $b$ boundaries
corresponding to D3- and/or D7-branes. We will attach $M$ 
external lines to the outer boundary as depicted in Fig.1.
We need to sum over all possible twisted boundary conditions for the boundaries.
The boundary conditions must satisfy the requirement that
\begin{equation}\label{mono1}
 \gamma^{\mu_1}_{a_1}=\prod_{s=2}^{b} \gamma^{\mu_s}_{a_s}~,
\end{equation}
where $\gamma^{\mu_1}_{a_1}$ corresponds to the outer boundary, and 
$\gamma^{\mu_s}_{a_s}$ ($s=2,\dots,b$) 
correspond to the inner boundaries.

{}Let us first consider the ${\cal N}=4$ theories for which the orbifold group
$\Gamma$ is trivial. (Note that in this case we can only have D3-branes
as introduction of D7-branes would break some number of supersymmetries.) 
The computation of any correlation function in the orientifold
theory (with $SO(N)$ or $Sp(N)$ gauge group) is reduced to the corresponding
computation in the oriented ${\cal N}=4$ theory (with $U(N)$ gauge group) up to factors
of $1/{2}$ (coming from the difference in normalizations of the corresponding D-brane boundary states in the oriented and unoriented cases). Such a simplification is due to
the fact that the unoriented world-sheets with cross-caps give contributions suppressed
by extra powers of $1/N$. 

{}Next, consider cases where $\Gamma$ is non-trivial (and hence supersymmetry is reduced).
If all the twisted boundary conditions are trivial ({\em i.e.}, $a_s=1$ for all $s=1,\dots,b$)
then the corresponding amplitude is the same as in the 
${\cal N} =4$ case (modulo factors of $1/{|\Gamma|}$).
Therefore, such amplitudes do not contribute to the gauge
coupling running (for which we would have $M=2$ gauge bosons attached to
the outer boundary) since the latter is not renormalized in ${\cal N}=4$ gauge 
theories due to supersymmetry. 

{}Let us now consider contributions with non-trivial twisted
boundary conditions. Let $\lambda_r$, $r=1\dots M$, be the Chan-Paton matrices corresponding to the
external lines. Then the planar diagram with $b$ boundaries has the following
Chan-Paton group-theoretic dependence:
\begin{equation}
 \sum {\rm Tr}\left(\gamma^{\mu_1}_{a_1} \lambda_1\dots\lambda_M\right)
\prod_{s=2}^{b}  {\rm Tr}(\gamma^{\mu_s}_{a_s})~,
\end{equation}
where the sum involves all possible distributions of $\gamma_{a_s}$ twists
(that satisfy the condition (\ref{mono})) as well as permutations of $\lambda_r$
factors. If the condition (\ref{Klein}) is satisfied, {\em i.e.}, if all the twisted Chan-Paton 
matrices are traceless, then the situation is analogous to that in the
oriented cases. That is, the only planar diagrams that
contribute are those with trivial boundary conditions. 
Such diagrams with all the boundaries corresponding to D3-branes (up to numerical
factors) are the same as in the parent ${\cal N}=4$ gauge theory.
The diagrams with trivial boundary conditions but with some boundaries corresponding
to D7-branes are subleading in the large $N$ limit as the numbers of D7-branes
(that is, the traces
${\mbox{Tr}}(\gamma^\mu_1)$ corresponding to D7-branes) are of order one.
This establishes that computation of any $M$-point correlation faction in the
large $N$ limit reduces to the corresponding ${\cal N}=4$ calculation in {\em oriented} 
theory, and that
these gauge theories are superconformal in this limit\footnote{Just as in
\cite{BKV}, it is also straightforward to show that 
a large class of non-planar diagrams without handles and cross-caps also vanish.}.

{}Now consider the cases where some of the twisted Chan-Paton matrices are {\em not}
traceless. Then there are going to be corrections to the $M$-point correlators coming
from planar diagrams with non-trivial twisted boundary conditions. These diagrams are 
subleading in the large $N$ limit as the corresponding traces are always of order one. 
This follows from the tadpole cancellation conditions (\ref{tad}) where the coefficients
$B_a$ and $C^\mu_a$ are of order one, so for $a\not=1$ we have ${\mbox{Tr}}(\gamma^\mu_a)\sim 1$. This implies that even for non-finite theories
computation of the correlation functions reduces to the corresponding computation in the
parent ${\cal N}=4$ oriented theory. 

{}Here we should point out that ``non-finiteness'' of such theories is a subleading 
effect in the large $N$ limit. This is because the $\beta$-function coefficients grow
as
\begin{equation}\label{beta}
 b_s=O(N^s)~,~~~s=0,1,\dots~,
\end{equation}
instead of $b_s=O(N^{s+1})$ (as in, say, pure $SU(N)$ gauge theory). This can be seen
by considering planar diagrams with $M=2$ external lines corresponding to gauge bosons.
Note that in string theory running of the gauge couplings in the low energy effective
field theory is due to infrared divergences corresponding to massless modes \cite{kap}.
The diagrams with all the boundaries corresponding to D3-branes and with all the 
boundary conditions corresponding to the identity element of $\Gamma$ are the same
(up to overall numerical factors) as in the parent ${\cal N}=4$ theory. Such diagrams,
therefore, do not contain infrared divergencies, and thus do not contribute the gauge 
coupling running. (That is, their contributions to the $\beta$-function coefficients $b_s$
vanish.) Therefore, the only diagrams that can contribute to the $\beta$-function coefficients $b_s$ are those with some boundaries corresponding to D7-branes and/or having twisted boundary
conditions with ${\mbox{Tr}}(\gamma^\mu_a)\not=0$. These are, however,
suppressed at least by one power of $N$ since the numbers of D7-branes are of order one,
and also such ${\mbox{Tr}}(\gamma^\mu_a)\sim 1$. This establishes (\ref{beta}).

{}Note that the estimates for the $\beta$-function coefficients in (\ref{beta}) for $b_{s>0}$
are non-trivial from the field theory point of view as they imply non-trivial cancellations
between couplings (such as Yukawas) in the gauge theory. On the other hand, within 
string perturbation expansion these statements become obvious once we carefully 
consider twisted boundary conditions and tadpole cancellation.

\section{Comments}

{}Here we would like to comment on some issues concerning the discussion 
in the previous sections.

{}Note that the entire argument in the previous section crucially depends
on the assumption that there is a well defined world-sheet description of the
orientifold theories at hand. Naively, it might seem that orientifolds of Type IIB
on ${\bf C}^3/\Gamma$ should have such world-sheet descriptions for any
$\Gamma$ which is a subgroup of $Spin(6)$. This is, however, not the case 
\cite{KST}\footnote{Some examples of this were also given in \cite{Zwart}.}.
In particular, there are certain cases where perturbative description is
inadequate as there are additional states present in the corresponding orientifolds
such that they are non-perturbative from the orientifold viewpoint \cite{KST}.
These states can be thought of as arising from D-branes wrapping various (collapsed) 
2-cycles in the orbifold. This leads to a rather limited number of large $N$ gauge theories
from orientifolds that have world-sheet description. 

{}The ${\cal N}=2$ supersymmetric large $N$ gauge theories from orientifolds were 
constructed in \cite{zura}. (The ${\cal N}=2$ theories with just orientifold 7-planes but no orientifold 3-planes were studied in \cite{FS}). The ${\cal N}=1$ theories based on Abelian orbifolds were constructed in \cite{zura,zura1} (some of these theories were subsequently discussed in \cite{blum}), and generalization to a non-Abelian orbifold
was given in \cite{zura2}. Finally, ${\cal N}=0$ orientifold theories were constructed in \cite{zura3}. As we already mentioned, the number of large $N$ gauge theories from orientifolds is quite limited. This fact is a bit mysterious from the field theory viewpoint. However, if one tries to find consistent models within the field theory framework, one ends up with the answer which completely agrees with the string theory predictions \cite{zura1}. In string theory, on the other hand, one can understand why such theories are so constrained from the fact that most of the orientifolds do not possess a world-sheet description \cite{KST}. Conversely, construction of large $N$ gauge theories from orientifolds, which are in agreement with {\em a priori} independent field theory expectations, is an important non-trivial check \cite{zura1} for correctness of the corresponding conclusions in \cite{KST}. This, in particular, has led to further developments in understanding of four dimensional Type I compactifications \cite{typeI}.

\end{document}